\renewcommand\vec{\boldsymbol{##1}}

\documentclass[prx,aps,twocolumn,amsmath,amssymb,floatfix,superscriptaddress]{revtex4-2}
\usepackage{physics}
\usepackage{amsmath}
\usepackage[dvips]{graphics}
\usepackage{bm}
\usepackage{xcolor}
\usepackage{appendix}
\usepackage{graphicx}
\usepackage{dcolumn}
\usepackage{bm}
\usepackage[colorlinks=true, allcolors=blue]{hyperref}

\newcommand{\jmq}[1]{\textcolor{red}{#1}}
\newcommand{\jm}[1]{\textcolor{gray}{#1}}

\def\vec#1{\mathbf{#1}}

\begin{document}

\preprint{APS/123-QED}

\title{Driven-dissipative fermionized topological phases of strongly interacting bosons  }

\author{Arkajyoti Maity }
\thanks{maity@pks.mpg.de}
 

\affiliation{Max Planck Institut fur Physik komplexer Systeme,
\\Nöthnitzer Str. 38, Dresden, 01187, Germany}

\author{Bimalendu Deb}
 
\affiliation{
Indian Association for the Cultivation of Science, \\ 2a and 2b Raja S.C. Malllik Road, Jadavpur, Kolkata 700032, West Bengal, India}

\author{Jan-Michael Rost$^{1}$}

\date{\today}

\begin{abstract}
 We study the optical response of a one-dimensional array of strongly nonlinear optical microcavities with alternating tunnel transmissivities, mimicking the paradigmatic Su-Schriefer Heeger model. We show that the non-equilibrium steady state of the bosonic system contains clear signatures of fermionization when the intra-cavity Kerr non-linearity is stronger than both losses and inter-site tunnel coupling.  
 Furthermore, changing the experimentally controllable parameters  
 detuning and driving strength, 
 in a topologically non-trivial phase, one can selectively excite either the bulk or edge modes or both modes, revealing interesting topological properties in a non-equilibrium system.
 
\end{abstract}

 \maketitle



 %
%

The physics of symmetry-protected topological phases (SPT) has been a topic of intense research over the last decade~\cite{hasan2010colloquium,qi2011topological,senthil2015symmetry}. Topologically non-trivial phases of matter were initially predicted and classified for non-interacting systems~\cite{Classification1}, with the appearance of robust edge states serving as the primary experimental evidence. First observed ~\cite{xu2014observation,hsieh2009tunable} in insulating materials, topological physics has since been realized with artificial designer lattices in mechanical~\cite{mechtopo,ma2019topological}, ultracold atomic~\cite{cooper2019topological,goldman2016topological}, photonic~\cite{lu2014topological,ozawa2019topological} and polaritonic~\cite{klembt2018exciton,solnyshkov2021microcavity} settings, unravelling novel SPTs, edge states involving synthetic dimensions~\cite{ozawa2019topological}, quasi-crystalline order~\cite{kraus2012topological,maity2021engineering} and much more. Interestingly and contrary to initial expectations, studies with moderate interactions between particles  have led to several novel classifications of SPTs~\cite {wang2014classification,rachel2018interacting}.
 \begin{figure}
 \centering
\includegraphics[width=1\columnwidth]{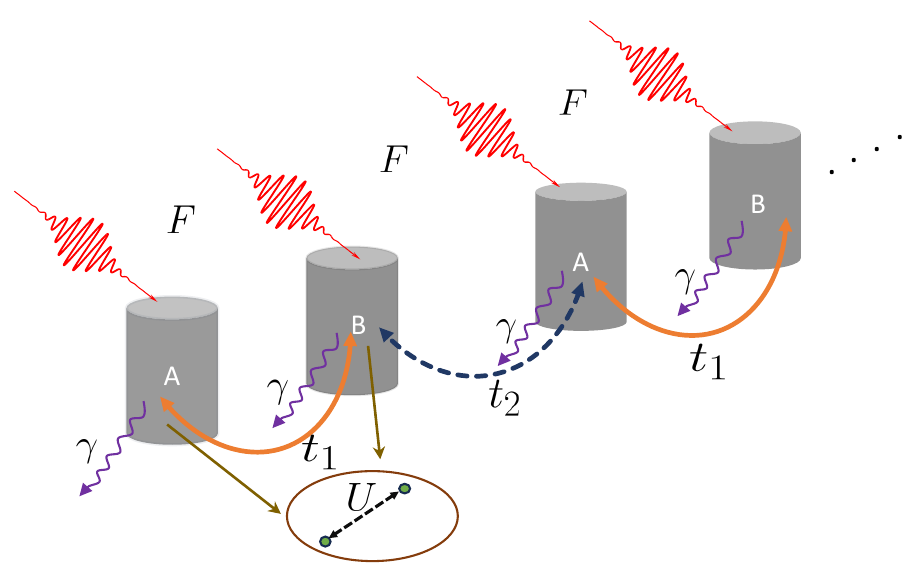}
\caption{Schematic diagram of our proposed set-up, featuring an array of non-linear semiconductor microcavities with alternating inter-pillar spacings for mimicking an SSH model. The cavities are driven homogeneously by an external laser field $F$. }
\label{schematic}
\end{figure}
In photonic platforms, nonlinear interactions have led to
topological phase transitions~\cite{maczewsky2020nonlinearity,hadad2018self}, mode-mixing ~\cite{mittal2018topological,kruk2019nonlinear}, topologically non-trivial gap-solitons~\cite{pernet2022gap,mukherjee2020observation,solnyshkov2017chirality,guo2020weakly} and the emission of symmetry-protected coherent light~\cite{st2017lasing,parto2018edge}. These studies, however, were performed for weak interactions, where mean field theories are applicable. In the strong interaction regime, more robust SPTs may arise. Examples are the observation of edge states in the many-body  Su-Schreifer-Heeger model, emulated with  Rydberg atoms~\cite{de2019observation} or the emergence of topological order~\cite{Wentopo}, such as the fractional quantum Hall phase~\cite{jain1990theory,stormer1999fractional}, recently realised, with artificial gauge fields in photonic setups~\cite{wang2024realization,knuppel2019nonlinear}. Photonic platforms also enable controlled engineering of gain and losses leading to the recent development of non-Hermitian topological phases~\cite{NHtopology,bergholtz2021exceptional,NESS1,ness2,NHBT,wanjura2020topological,gomez2023driven,Transient1,Transient2}. Applications include perfectly directional robust topological amplification and frequency conversion~\cite{wanjura2020topological,2phtndrive}.  
Topological photonics in such non-equilibrium open systems, coupled additionally with strong non-linearities is a nascent 
but promising field of research~\cite{Gauge,roccati2024hermitian,cao2024topological,kawabata2022many}.\\
Here, we identify topological signatures in a non-equilibrium state of strongly correlated bosons and demonstrate how they can be manipulated by changing experimentally accessible parameters. The state is created in a strongly interacting Su-Schriefer-Heeger(SSH) lattice~\cite{SSH} of optical micro-cavities mutually coupled by tunnelling and driven by a coherent laser field. We specifically focus on the strongly nonlinear regime, where nontrivial quantum correlations between bosons, constrained in one spatial dimension, herald the onset of a Tonks-Girardeau gas ~\cite{girardeau1960relationship,paredes2004tonks,kinoshita2004observation} of 'fermionized bosons'. Variants of the SSH model, including particle-particle interactions, driving and losses have been explored over the years leading to several theoretical proposals~\cite{azcona2021doublons,matveeva2024weakly, Faugno,stepanenko2020interaction} and experimental realisations~\cite{st2017lasing,downing2019topological,parto2018edge,pernet2022gap}. However, to the best of our knowledge,  the interplay between strong correlation-induced fermionization and topology in an interacting non-equilibrium SSH system has not been studied. Indeed, we observe that the reduced spatial dimensions of interacting bosons, in conjunction with drive and dissipation can produce highly non-trivial 'fermionized' steady states, where one can selectively excite edge modes, bulk modes or both by changing the driving parameters, without having to resort to any predefined state preparation.
To be specific, we consider a coupled one-dimensional array of $M$ optical cavities embedded in a non-linear semiconductor medium~\cite{scmicro1,scmicro2}, with each cavity being subjected to uniform driving, as sketched in Fig.~\ref{schematic}. Cavity polaritons, which are composite bosons originating from the strong coupling between semiconductor quantum well excitons and cavity photons are particularly well suited for our proposal and have already shown great promise in emulating interacting non-Hermitian topological lattice models ~\cite{solnyshkov2021microcavity, xu2022non,szameit2024discrete,pernet2022gap,mandal2020nonreciprocal,zhao2024observation}. Their excitonic component can give rise to a sizeable Kerr nonlinearity~\cite{carusotto2013quantum,chang2014quantum,cavitypolaritons1,cavitypolaritons2}, while their photonic component allows for engineering loss and gain in the system. In our model,  we restrict the physical description to a single mode of frequency {$\omega_{0}$} per cavity by ensuring that the energy spacing between neighbouring modes of a single cavity is much larger than all other energy scales. The Hamiltonian describes an SSH Bose-Hubbard model, where the Hubbard interaction term $U$ arises due to the Kerr non-linearity~\cite{chang2014quantum} of the intra-cavity medium. Alternating couplings ({$t_{1}$} and {$t_{2}$}) are created by engineering staggered overlap between nearest neighbour cavity modes~\cite{pernet2022gap,solnyshkov2021microcavity}. Driving every cavity uniformly with  coherent monochromatic light of  amplitude {$F$} and frequency {$\omega_{p}$}, we recast the Hamiltonian in a frame rotating with the drive~\cite{carusotto2013quantum} as 
$ H_{\rm tot}=H_{\rm sys}+H_{\rm drive} $, where  
\begin{equation}
\begin{gathered} \label{eq:1}
H_{\rm sys} = \hbar{\Delta\omega}\sum_{j} (a^{\dagger}_{j}a_{j}+ b^{\dagger}_{j}b_{j})- 
{t_{1}}\sum_{j}(a^{\dagger}_{j}b_{j}+h.c)
\\  -{t_{2}}\sum_{j}(b^{\dagger}_{j}a_{j+1} +h.c) +{U}\sum_{j}(a^{\dagger}_{j}a^{\dagger}_{j}a_{j}a_{j}+b^{\dagger}_{j}b^{\dagger}_{j}b_{j}b_{j} )\,
\end{gathered}
\end{equation}
 with the detuning ${\Delta\omega=\omega_{0}-\omega_{p}}$ of the cavity frequency from the pump frequency.
The one-photon resonant drive is described by
\begin{equation}
\begin{gathered} \label{eq:2}
H_{\rm drive} = \sum_{j}{F} (a^{\dagger}_{j}+b^{\dagger}_{j})+h.c.\,,
\end{gathered}
\end{equation}
where $a_{j} (b_{j})$ denotes bosonic annihilation operators for sublattice A (B) in unit cell $j$, as is the case in the SSH model (see Fig.\ref{schematic}). 

To compute the steady state,  under particle loss  from the system at a dissipation rate $\gamma$, we assume Born-Markov conditions and make use of the Lindblad master equation~\cite{breuer2002theory},
\begin{equation}
\label{eq:LB}
\partial{\rho}/{\partial t}=-i[H,\rho]+D[\rho]
\end{equation}
with
\begin{equation}
\begin{split}
\begin{gathered} \label{eq:3}
D[\rho]=\frac{\gamma}{2}\sum_{j}(2a_{j}\rho a^{\dagger}_{j}-a^{\dagger}_{j}a_{j}\rho -\rho a^{\dagger}_{j}a_{j}) \\
 + \frac{\gamma}{2}\sum_{j}(2b_{j}\rho b^{\dagger}_{j} - b^{\dagger}_{j}b_{j}\rho -\rho b^{\dagger}_{j}b_{j})\,.
\end{gathered}
\end{split}
\end{equation}

\begin{figure}

\centering

\includegraphics[width=0.85\columnwidth]{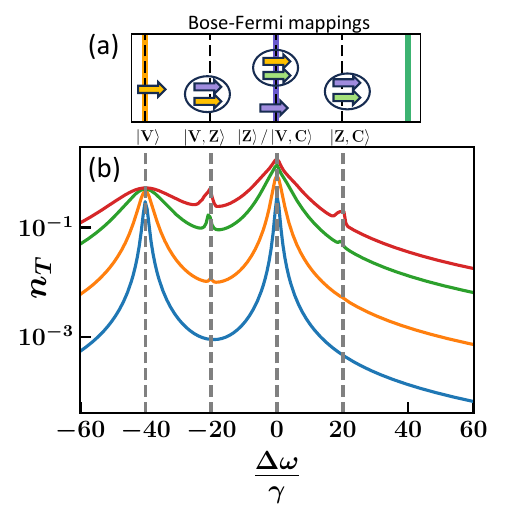}
\caption{(b) Transmission spectrum of $M=4$ cavities in the impenetrable boson limit, with $t_1/\gamma{=}2, t_2/\gamma{=}{4}0$ in Eq.~\eqref{eq:1} computed from the
steady state $\partial\rho/\partial t = 0$ in Eq.~\eqref{eq:LB}. The different curves correspond from below to increasing pump amplitudes $F/\gamma = 0.1,1.0,3.0,5.0$. The vertical dashed lines indicate the detunings for the spectral peaks, derived analytically from the eigenvalues of the Fermi-mapping. Each mapping is sketched in (a) with the colored arrows representing the different orbital contributions, either from the bulk (orange ($\ket{\rm{V}}$) and green ($\ket{\rm{C}}$), or the topological edge (light purple, $\ket{\rm{Z}}$). All two-particle mappings are circled. }
\label{fig:nearfieldspectrum}
\end{figure}

Since in most experimentally realizable setups, the propagation of the pump light through the sample is inhibited ~\cite{el2006polariton,amorev}, properties of the system are inferred from the spectrum and coherence of the transmitted light. The total transmitted intensity is
proportional to the average total number of particles, $ n_{T}=\left<\sum_{j}(a^{\dagger}_{j}a_{j}+ b^{\dagger}_{j}b_{j})\right>$, in the
steady state~\cite{pernet2022gap}, while the near-field transmission \cite{pernet2022gap,nearfield} corresponds to the site resolved population. 

For better understanding, we briefly line out the most important features of the eigenstates of any one-dimensional system of \textbf{N}  impenetrable bosons ($U{\rightarrow} \infty$) in the absence of pumping and dissipation. A more detailed analysis is given in Appendix~\ref{appendix:a}.
The extreme correlation-induced fermionization of bosons implies that strongly interacting bosons can be mapped to a non-interacting fermionic system. The spatial eigenfunction, $\psi_{B}(\mathbf x)$ of \textbf{N} bosons at positions $\mathbf x = (x_1,\ldots,x_N)^\dagger$ can be mapped to a fermionic one $\psi_{F}(\mathbf{x})$, as $ \psi_{B}(\mathbf x)=|\psi_{F}(\mathbf{x})|$. The fermionic wavefunction can then be classified by the occupation numbers of single-particle orbitals~\cite{girardeau1960relationship}.  One can invoke the coordinate Bethe ansatz ~\cite{lieb1963exact,lieb2}
to write  $\psi_{B}(\mathbf x)$ as
\begin{equation}
\label{eq:fermionise}
\psi_{B}(\mathbf x) \sim \det[\Phi_{j}(x_{l})] \prod_{1\leq j<l\leq N}\!\!\!\!\!{\rm sgn}(x_{j}-x_{l})\,.
\end{equation}
The number of permitted single-particle orbitals $\Phi_{j}(x_{l})$, 
depend on the degrees of freedom of the system and 
the sign function is multiplied with the Slater determinant to fix the particle statistics from fermionic to bosonic. The total energy $E_{N}$ of the state is then exactly replicated by the sum of the  energies $\epsilon(j_{i})$ of the filled single-particle orbitals,
\begin{equation}
\label{eq:energy balance}
E_{N}=\sum_{i=1}^{N}{\epsilon(j_{i})}\,,
\end{equation}
where  $j_{i}$ denotes the orbital occupied by the $i$'th mapped particle. Subjecting such a bosonic system to external driving and dissipation, especially in lattice environments can create very non-trivial Bose-Fermi mappings in the steady state as demonstrated for the case of a simple 1D chain ~\cite{Ciuti}. We focus on investigating the interplay of fermionization and topology in the driven-dissipative SSH Bose-Hubbard model described by \eqref{eq:1} for both periodic and open boundary conditions. Although imposing periodic boundary conditions leads to interesting features of quasi-momentum quantized mapping and non-trivial band couplings, we leave its discussions to Appendix~\ref{appendix:b} and focus here on the open boundary case, where topological effects appear physically in the form of edge states via the bulk-boundary correspondence~\cite{chen2020elementary}. 
\begin{figure}[h!]
\centering
\includegraphics[width=1\columnwidth]{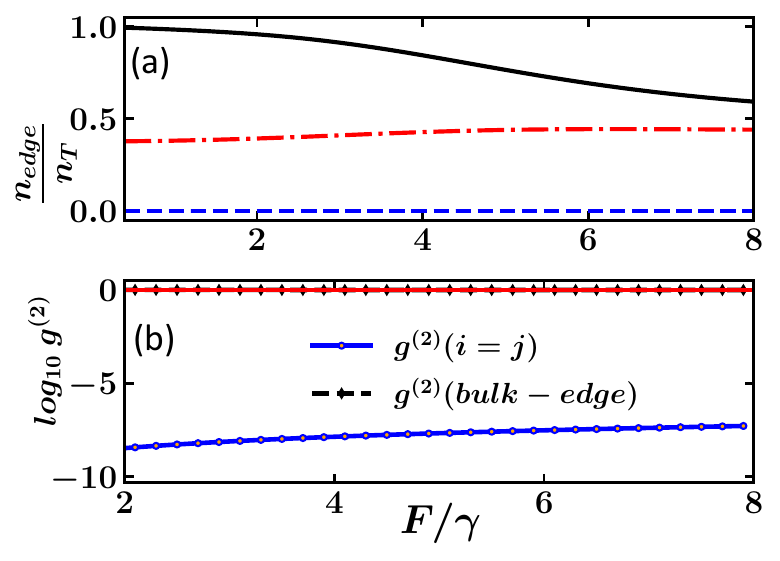}

\caption{The edge-population fraction, plotted against the drive amplitude $F$ is shown in (a), with the cavity-pump detuning ($\Delta\omega/\gamma$) locked in at three distinct analytically predicted fermionized resonances: $\ket{\rm{V}}$(blue dashed), $\ket{\rm{\rm{V,Z}}}$ (red dash-dotted) and $\Delta\omega/\gamma=0$ (black solid), corresponding to either $\ket{\rm{Z}}$ or $\ket{\textbf{\rm{V,C}}}$
(b) Plot of the lattice averaged zero time auto ($i=j$) and bulk-edge intensity cross-correlations against $F/\gamma$ for $\Delta\omega=0$ in the driving regime where both bulk and edge are populated. 
The orange solid line is indicative of non-correlated Poissonian photon emission statistics.}
\label{Fig_g^2edge}
\end{figure}
We numerically evaluate the steady state density matrix from \eqref{eq:LB} for $M=4$ cavities in the topologically non-trivial open boundary configuration  $U {\rightarrow }\infty$ and  $t_{1}{ <} t_{2}$ respecting  the Born-Markov regime of weak system-reservoir coupling ($\gamma {\ll} t_{2} {\ll} U{\rightarrow} \infty $) by choosing $t_1/\gamma{=}2$ and $t_2/\gamma{=}{40}$. The expected transmission (obtained from $n_{T}$) is shown in Fig.~\ref{fig:nearfieldspectrum}, as a function of the cavity-pump detuning ($\Delta \omega$) for a range of drive amplitudes $F$. The Bloch state quasi-momenta $q$ (defined in Appendix.\ref{appendix:b}) are no longer good quantum numbers for open boundary conditions. Hence, the single particle orbital energies ($\xi$) are obtained by exact diagonalization of \eqref{eq:1} with $U=0$. 
 Fixing the value of $\gamma$ at unity, the resonant detuning at which the pump can excite any allowed \textbf{N}-boson excited state is now expected at $\Delta \omega = E_{\rm total}/N =\sum_{i=1}^{N}\xi_{i}/N $. Fig.~\ref{fig:nearfieldspectrum} illustrates how our numerically determined mean particle number $n_T$ exhibits peaks at the analytical Bose-to-Fermi mapped states. For small $F$, the pump  excites an $N=1$ bosonic state giving rise to   single mapped fermionic states, namely $|\textup{V}\rangle$ from the bulk at detuning
$\Delta\omega\approx -40$ and $|\rm{Z}\rangle$ from the topological edge at $\Delta\omega=0$, while for increasing field strength $F$,  $N>1$ bosonic states are created through multi-photon processes, giving rise to additional two fermion peaks (indicated with circles in the sketch of mapped states Fig.~\ref{fig:nearfieldspectrum}a)  appearing near $\Delta\omega =\pm 20$.

The zero detuning case is fascinating because it can realize two different mappings. The first mapping $|\rm{Z}\rangle$, already mentioned, corresponds to many-body bosonic states represented by zero-energy single-particle states which are the degenerate topological gapless edge modes. The second mapping corresponds to a $\textbf{N}=2$ excited state $\ket{\rm{\rm{V,C}}}$, with one mapped fermion in $\ket{\rm{V}}$ and the other 
in its chiral partner, the upper energy state, $\ket{\rm{C}}$, such that their energies sum up to zero. Note, that the mapping $\ket{\rm{C}}$ by itself cannot be excited with a single photon due to symmetry constraints~\cite{supp}. 

 Unlike $\ket{\rm{Z}}$, the two-particle state $\ket{\rm{\rm{V,C}}}$ is completely immersed in the lattice bulk. The results of Fig.~\ref{fig:nearfieldspectrum} suggest that the population of edge versus bulk states changes with increasing drive strength $F$ differently, for the fermionized resonances which would allow the selective population of bulk and edge modes. This is indeed the case,
 as shown in Fig.~\ref{Fig_g^2edge}(a).  When pumping at zero detuning with small $F$, 
 exclusively edge states are populated (black line), while pumping at detuning $\Delta\omega$, corresponding to the fermionized resonance $|\rm{V}\rangle$ (blue line), only bulk modes are excited.
 Not surprisingly,  the fraction of edge states of the double excitation $|\rm{V,Z}\rangle$ (red line) comprised of a bulk and an edge state, is hardly sensitive to the drive strength.
 Suppose one is interested in switching between edge and bulk excitation. This can be achieved at zero detuning by simply increasing $F$, since small $F$ will favour the singly excited edge state $|\rm{Z}\rangle$, while larger $F$ will also populate the doubly excited bulk state $|\rm{V,C}\rangle$, (see black line).
 
 To gain further insight into the microscopic nature of the many-body states, we investigate the intensity correlations in the near-field transmission. It is well known that although the density distributions are identical, the one-dimensional hard core bosonic wavefunction is not the same as the non-interacting fermionic case. This can be verified by calculations of the momentum distribution in periodic boundary conditions or by probing any odd-order correlation function. For example, the momentum distribution of a Tonks gas is very different to that of free fermions.~\cite{Pezer, Scopa,paredes2004tonks}. Any even-order correlation should however be identical because it depends on field densities ~\cite{astrakharchik2005beyond,hao2022n,girardeau2001quantum}.
 
 Having already established that our system's steady state can host both topological fermionized edge state(s) and bulk states at zero detuning, we investigate how the bosonic modes on the bulk and edge are correlated to each other. To do so, we calculate the equal time second-order cross-correlation function between lattice sites $i$ and $j$, defined as
\begin{equation}
\label{eq:g2}
   g^{(2)}(i,j)=\frac{\left<{c_{i}}^\dagger {c_{j}}^\dagger c_{j}c_{i} \right>}{\left<{c_{i}}^\dagger c_{i}\right>\left<{c_{j}}^\dagger c_{j}\right>}\,,
\end{equation}
where $c_{j}$  is the boson annihilation of site $j$, belonging  to sublattice A(B) depending on whether $j$ is odd(even) and the expectation value is calculated at steady state. Fig.~\ref{Fig_g^2edge}(b) shows the lattice averaged on-site and bulk-edge correlations for the topological configuration at zero detuning  $\Delta \omega{=}0$. One can observe that the on-site ($i{=}j$) correlations are extremely small, indicating strong anti-bunching in the transmission statistics. This is to be expected since the repulsive interaction effectively fermionizes the system, such that a single site cannot host more than one particle:  the mapping to fermions implies Pauli blocking of two similar fermions at the same spatial coordinate. This is analogous to strong non-linearity induced photon blockade~\cite{Ciuti,pb&j}, used to generate single photons. Interestingly, the bulk-edge correlations of the transmission are fully separable, being coherent for a large range of drive parameters. This can be explained by analysing the drive-controlled excitation, as the inherent fermionization governs the intensity correlations. For drive amplitudes $F$, where the steady state can host fermionized bulk modes $\ket{\rm{V,C}}$, the correlation between the edge $\ket{\rm{Z}}$ and the bulk gets suppressed due  to the topological protection of the edge states imposed by chiral symmetry~\cite{asboth2016short}. This means that each cavity can act as a single photon source, tunable from  emitting solely from the edges to emitting from the bulk and the edge by increasing the driving strength at zero detuning. However, since the emission statistics are uncorrelated,  any perturbation in the bulk will not disturb the emission from the edges, as long as the symmetries protecting the phase~\cite{asboth2016short,wang2023sub} are maintained.
\begin{figure}[h!]
\centering
\includegraphics[ width = 1.01\columnwidth]{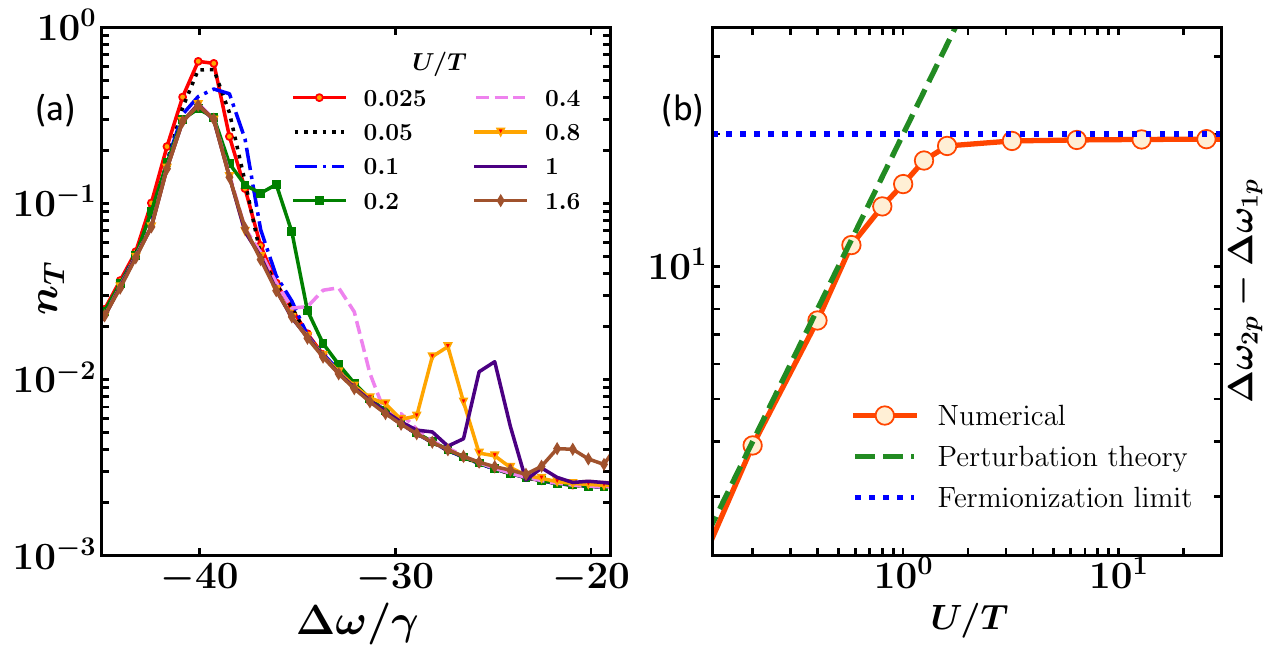}
\caption{The lowest energy one boson and two boson peaks in the transmission spectrum as calculated from the steady state are shown in (a) for a range of interaction strengths $U/T$. Panel (b) shows the shift in resonant detunings for the two peaks against the interaction strength. For $U/T \approx 2$, the shift saturates to the predicted shifts of the fermionized mapped states (blue dotted line). The green dashed line shows the linear variation of the shifts against $U/T$, as predicted by perturbation theory.}
\label{fig:shifts}
\end{figure}
We conclude our discussion on these fermionized topological phases by assessing for which parameter regime one can observe fermionization experimentally. A crucial question for experiments is the effective interaction strength $U/T$, $T=\max({t_1,t_2})$ (the largest kinetic energy scales in the system)  at which the bosonic system behaves as its fermionized counterpart.
To this end, we analyze the excitation dynamics as a function of $U/T$ for fixed driving strength $F/\gamma {=} 1$.
Fig.~\ref{fig:shifts}(a) shows the total transmission $n_T$ versus the detuning $\Delta\omega/\gamma$.
The spectrum shows a strong peak at $\Delta\omega_{1p}$, corresponding to the one-boson excitation
and a second peak of the two-boson excitation, whose position $\Delta\omega_{2p}$ gets increasingly blue shifted from $\Delta\omega_{1p}$ for larger ${U}/{T}$. Eventually, $\Delta\omega_{2p}$ reaches an asymptotic value (seen in  Fig.~\ref{fig:shifts}(b)), where the two-boson excitation coincides in the steady state with the 'fermionized' 2 particle peak $\ket{\rm{V, Z}}$, which is roughly the case at $U/T \approx 2$.  The behaviour can be understood using perturbation theory.
%
In the $U{=}0$ limit, a two-particle bosonic state will have both particles occupying $\psi_{V}$, the wavefunction of the lowest energy single body orbital $\ket{\rm{V}}$. The zeroth order wavefunction can be written as $\psi_{0}=\psi_{V}\otimes \psi_{V}$. The resonant detunings at $U=0$, given by \eqref{eq:energy balance}, for the one-particle ($\Delta\omega_{1p}$) and two-particle ($\Delta\omega_{2p}$) states coincide since the energy of the one-boson state is just twice the energy of the two-boson state. Increasing $U$, the energy gets shifted perturbatively by
$\Delta E_{0} = U\bra{{\psi_{0}}}\hat{\alpha}\ket{\psi_{0}}$
with 
$\hat{\alpha}=\sum_{i}(a^{\dagger}_{i}a^{\dagger}_{i}a_{i}a_{i}{+}b^{\dagger}_{i}b^{\dagger}_{i}b_{i}b_{i} )$.
%
Since the two-particle ground state wavefunction $\psi_{0}$, for our system, is a spatially symmetric superposition of the bulk sites, we have $\Delta E_{0} \approx  U$. 
Fig.~\ref{fig:shifts}(b) reveals a linear increase of the shift $\Delta\omega_{2p}-\Delta\omega_{1p}$ in the interaction regime  of perturbatively small $U/T$, eventually 
crossing over to the fermionization regime at large $U/T$.


In conclusion, we have provided a scenario where the steady state of a many-body system can be steered from bulk to edge modes by just changing the strength of the optical drive. This has been enabled by  the interplay of topology and strong correlation-induced fermionization of bosons. As a consequence, we could show how to control the transmission intensity and statistics from the bulk and edge modes without resorting to prior state preparation. Finally and importantly, we have demonstrated that the predicted phenomena occur at parameters which can be realized in state-of-the-art photonic/polaritonic platorms~\cite{solnyshkov2021microcavity,pernet2022gap,bloch2022non,zhao2024observation}\\
\textit{Acknowledgements:} A.M would like to thank Kingshuk Adhikary and Panos Giannakeas for helpful discussions.  The authors would also like to acknowledge the open-source package QuTiP~\cite{johansson2012qutip}, used in some numerical calculations.


\appendix

\section{ Bose-Fermi mapping in a 1D hard-core continuum Bose gas}
\label{appendix:a}

The  natural starting point for a Bose-Fermi mapping of an impenetrable one-dimensional gas of bosons with  contact interaction $U{\to}\infty$ is  the coordinate Bethe ansatz \eqref{eq:fermionise}.   In equilibrium, this gas, known as the Lieb-Liniger model~\cite{lieb2}, is a rare, fully integrable system. To evaluate the eigenstates with periodic boundary conditions one first inserts  plane waves  into the Bethe ansatz,
$\Phi_j(x_l)=e^{ik_jx_l}$;
a rigorous treatment can be found in Refs.~\cite{lieb1963exact,lieb2}.
For $U{\to}\infty$,  known as the Tonks-Girardeau limit, the eigen solutions for  $N$ bosons get fermionized. The wavefunction  for bosons at positions $\mathbf x = (x_1,\ldots,x_N)^\dagger$, $\chi_{N}$ is given by
\begin{equation}
\label{eq:A1}
\chi_{N}(\mathbf x)=\frac{C}{\sqrt{N!}}\det\left[e^{ik_{j}x_{l}}\right]\prod_{1\leq l<j\leq N}\!\!\!\!\!{\mathrm {sgn}(x_{j}-x_{l})}
\end{equation}
with the quasi-momentum $k_{j}$ for the j'th boson and  normalization constant $C$. The construction ensures that 
$\chi_{N}(\mathbf x)$ vanishes for $x_{j}=x_{l}$. Furthermore, the total momentum of the eigenstate is given by $P_{N}=\sum_{j}{k_{j}}$ and the total energy by $E_{N}=\sum_{j}{k_{j}^2/2}$ ($\hbar=1$). Note, however, that the quasi momenta do not represent the true momentum of the bosons, although their sums are necessarily equal~\cite{yukalov2005fermi}. Ensuring periodic boundary conditions by confining our system on a ring of length $L$, the wavefunctions  \eqref{eq:A1}  must necessarily satisfy
$\chi_{N}(\vec x)|_{x_i=0}=\chi_{N}(\vec x)|_{x_i=L}$  for all $i {=} 1,\ldots, N$.
With \eqref{eq:A1} it is easy to see that this requires quantising the quasimomenta,
%
\begin{align}
\label{eq:A2}
k_{i}& = &(n_{i}+1/2)&\,\,{2\pi}/{L}& & N=\rm{even},\nonumber\\
k_{i}& = &n_{i}&\,\,{2\pi}/{L}  &&N=\rm{odd}, 
\end{align}
where $n_{i}$ is an  integer.
This peculiar quantization of quasimomenta is one of the hallmarks of fermionization. 

\section{ Spectral signatures of fermionization in  periodic boundary conditions}
\label{appendix:b}

Having established in brief some salient points on the fermionization of the 1D hard-core Bose gas, we move on to tackling the driven dissipative hard-core SSH Bose-Hubbard model described in Eqns.~\eqref{eq:1} and \eqref{eq:2} with periodic boundary conditions imposed.
Since the SSH model hosts a periodic lattice potential, for
periodic boundary conditions, in the non-driven regime, any fermionized $N$ particle bosonic state can be classified via the Bloch states of the Hamiltonian, $\ket{q_{1}^{m_{1}}, q_{2}^{m_{2}},...,q_{N}^{m_{N}}}$, where  $m_{i}$ denotes the band index and $q_{i}$ denotes the crystal quasimomentum of the mapping for the $i$'th boson. An important point to note is that for a purely non-interacting bosonic system, there is no restriction in two or more particles occupying the same orbital($q_{i},m_{i}$), while in the strongly correlated regime,  fermionic states defined by the mapping  \eqref{eq:fermionise}  are Pauli blocked from having identical quantum numbers. Energy $E$ and  total momentum $P$
of the bosonic state are given by $E=\sum_{i}\epsilon(m_{i},q_{i})$ and $P=\sum_{i}q_{i}$, respectively, although the momentum distributions do not map~\cite{girardeau1960relationship,yukalov2005fermi}. When the drive is turned on, it can create non-trivial steady states, by generating excitations from the vacuum state to any allowed ${N}$ boson 'fermionized' state.

\begin{figure}

\centering

\includegraphics[width=1.0\columnwidth]{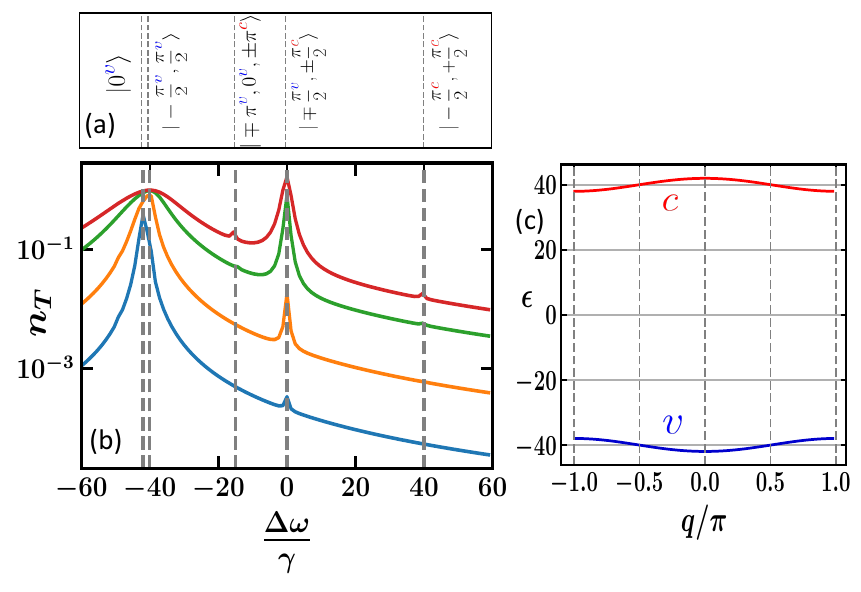}
\caption{(b) Transmission spectrum of a closed chain of $M{=}4$ cavities and $t_1/\gamma{=}2,t_2/\gamma{=}{4}0$ in \eqref{eq:1} computed from the
steady state, for $F/\gamma = 0.1$~({blue}), $1.0$ ~({orange}), $3.0$~({green}), $5.0$~ ({red}). (a) The detunings, corresponding to the fermionized resonances are indicated by dashed lines with each mapping being explicitly written out in terms of the quasimomenta and band contributions. (c) The band diagram of the SSH Hamiltonian for our chosen parameters is shown as a reference to the mappings depicted in (a).
}
\label{fig:nearfieldspectrum2}
\end{figure}

We observe this in Fig.~\ref{fig:nearfieldspectrum2}, where we plot the total transmitted intensity, numerically evaluated with \eqref{eq:LB} in the limit of impenetrable bosons and weak dissipation ($\gamma \ll \max(t_{1},t_{2})\ll U \to\infty $) for a range of drive strengths. We choose the lattice spacing between unit cells to be unity and the other parameters chosen are the same as those used in the main text. 
 Fig.\ref{fig:nearfieldspectrum2}(b) reveals resonances of the transmitted spectrum at specific detunings $\Delta \omega $. The position of these resonances does not depend on the driving strength, but their height sensitively depends on $F$. The positions are determined by the frequencies, at which the pump can excite any $N$ particle fermionised many-body bosonic eigenstate $\ket{q_{1}^{m_{1}}, q_{2}^{m_{2}},...,q_{N}^{m_{N}}}$ in the non equilibrium steady state and those frequencies analytically given by the total energy of the single particle orbitals normalised to the total number of particles,
\begin{equation}
\label{eq:Resonance1}
\Delta \omega = E_{\rm total}/N =\sum_{i=1}^{N}\epsilon_{m_{i}}(q_{i})/N\,.
\end{equation}
For small driving $F\ll\gamma$, the pump induces a steady state which has the largest overlap with the vacuum state and a single bosonic excitation $N=1$, which is Fermi-mapped to $\ket{0^{v}}$, a fermion in the lower Bloch band, denoted by $v$, at quasimomentum zero. For larger values of the drive amplitude, higher Fock states are excited (since the system has no U(1) symmetry), with the constraint  $\sum_{i}q_{i} = 0$ (mod $2\pi$), due to momentum conservation and the quantization condition for the quasimomenta, see \eqref{eq:A2} and \cite{Ciuti}.  Interestingly, in a two-band model like the SSH, the pump can also excite many-body bosonic states which map to fermionic excitations in two bands of the SSH model ($v$ and $c$), subject to the momentum conservation and the energy resonance relation \eqref{eq:Resonance1}. We find very good agreement of the numerically evaluated peak positions with that predicted by \eqref{eq:Resonance1} and list all the single-band and coupled-band mappings in Fig.~\ref{fig:nearfieldspectrum2}(a). The band diagram for the SSH model in our chosen parameters is provided in Fig.~\ref{fig:nearfieldspectrum2}(c) for reference. It should be noted that the transmission spectra upon exchanging $t_1$ and $t_2$ are identical. This is expected from their identical Bose-Fermi mappings due to the same equilibrium band structure. However, they are topologically not equivalent, and the implications of this fact are presented in the main text.

\bibliography{apssamp}

\begin{thebibliography}{87}
\bibitem [{sup()}]{supp}%
  \BibitemOpen
  \href@noop {} {}\bibinfo {note} {The mapping $\ket{\rm{C}}$ is the chiral partner of $\ket{\rm{V}}$ in the sublattice degrees of freedom (A/B); i.e the relative phase between the amplitudes of the wavefunction on sublattices A and B for both mappings differ by a factor of $\pi$. Hence, the mapping will be necessarily absent for $M = 4$, if the one-photon drive excites $\ket{\rm{V}}$. More formally, we do not observe any signal as $\bra{\rm{C}}H_{\textup{drive}}\ket{0} = 0 $, where $\ket{0}$ is the many-body vacuum state}\BibitemShut {NoStop}%
\end{thebibliography}

\end{document}